\documentclass[12pt]{article}
\usepackage{amsfonts}
\usepackage{amsmath}
\usepackage{geometry}

\input{tcilatex}

\begin{document}

\title{Modeling the phase behavior of polydisperse rigid rods with
attractive interactions, with applications to SWNTs in superacids}
\author{Micah J. Green$^{1}$\thanks{%
Current address: \ Department of Chemical Engineering, Texas Tech
University, Lubbock TX 79409, mgreen@alum.mit.edu}, A. Nicholas G.
Parra-Vasquez$^{1}$, \and Natnael Behabtu$^{1}$, Matteo Pasquali$^{1,2,\perp
}$ \\
$^{1}$Dept. of Chemical and Biomolecular Engineering, $^{2}$Dept. of
Chemistry\\
Rice University\\
6100 Main Street, MS362\\
Houston TX 77005\\
$^{\perp }$mp@rice.edu}
\maketitle

\begin{abstract}
The phase behavior of rodlike molecules with polydisperse length and
solvent-mediated attraction and repulsion is described by an extension of
the Onsager theory for rigid rods. A phenomenological square-well potential
is used to model these long-range interactions, and the model is used to
compute phase separation and length fractionation as a function of well
depth and rod concentration. \ The model closely captures experimental data
points for isotropic/liquid crystalline phase coexistence of single-walled
carbon nanotubes (SWNTs) in superacids. The model also predicts that the
isotropic-biphasic boundary approaches zero as acid strength diminishes,
with the possibility of coexistence of isotropic and liquid crystalline
phases at very low concentrations; this counterintuitive prediction is
confirmed experimentally. \ Experimental deviations from classical theories
for rodlike liquid crystals are explained in terms of polydispersity and the
balance between short range repulsion and long range attractions. \ The
predictions of the model also hold practical import for applications of
SWNT/superacid solutions, particularly in the processing of fibers and films
from liquid crystalline SWNT/superacid mixtures.
\end{abstract}

\section{Introduction}

When suspended in a liquid, rodlike molecules and particles transition from
an isotropic, disordered state to an aligned, liquid-crystalline nematic
state at sufficiently high concentration. \ This classical, peculiar
behavior has generated numerous theories for predicting the
isotropic-nematic transition and the equilibrium properties of the
coexisting phases; typically, these theories treat the rodlike molecule or
particle as an idealized Brownian rigid rod of high aspect ratio. \ These
theories capture with some success the behavior of classical lyotropic
liquid-crystalline materials such as liquid crystalline polymer solutions
and rodlike biological particles in water \cite{dogic}. \ However, previous
theoretical studies fail to describe dispersions of novel anisotropic
nanomaterials such as single-walled carbon nanotubes (SWNTs). \ Here we
extend the theoretical description of polydisperse Brownian rods to include
the competition of short-range repulsion and long-range attraction that is
known to yield rich behavior in colloidal phases. \ We focus on the
coexistence between the isotropic and nematic phases.

\subsection{Isotropic-Nematic Phase Separation}

In 1949, Onsager developed a statistical mechanical theory for the rod
orientation distribution function in order to describe the phase behavior of
idealized monodisperse rods interacting via a mean field excluded-volume
potential in an athermal solvent \cite{onsager}. \ This theory is based on
the competition between the orientational entropic contribution to the free
energy and the excluded-volume potential. \ This competition gives rise to
the isotropic-nematic phase transition that is characteristic of many
rodlike molecules. \ As rods crowd beyond a critical concentration,
alignment occurs spontaneously because orientational ordering is accompanied
by a loss of interaction free energy which offsets the partial gain of free
energy due to loss in orientational entropy. The predicted coexistence of an
isotropic phase and an aligned phase is observed in experimental systems. \
Onsager's original formulation was numerically refined by Kayser and Ravech%
\'{e} \cite{kayser} and Lekkerkerker \textit{et al}\ \cite{lekkerkerker}. \
Later studies extended Onsager's approach to nonhomogeneous systems in order
to resolve the profile of the interface between bulk isotropic and nematic
regions \cite{chennoolandi, kochharlen,shundyak} and finite-sized regions 
\cite{green1}. \ Onsager's approach was also modified to include attractive
interactions via a simple order-parameter-based attractive potential by
Khokhlov and Semenov \cite{LCinpolymers, semenov1985}; Khokhlov and Semenov
used the Parsons scaling approximation to allow Onsager's second virial
coefficient truncation to remain valid up to the high concentrations that
result from the attractive potential.

The major competing theoretical framework for rigid rod phase behavior is
Flory's lattice-based theory \cite{flory1956}, which accounts for
excluded-volume interactions via packing effects and allows for attractive
inter-particle interactions governed by the parameter $\chi $. \ Flory's
predictions for the boundaries of the well-known biphasic chimney for an
athermal solvent ($\chi =0$) are slightly higher ($7.89d/L$, $11.57d/L$)
than the Onsager predictions ($3.29d/L$, $4.19d/L$), where $d$ and $L$ are
the rod diameter and length, respectively. \ As the attractive force
increases, the biphasic region broadens and the concentration of the
coexisting aligned phase grows dramatically such that the coexistence
nematic phase is a solid. \ (The Onsager-based results of Khokhlov \cite%
{LCinpolymers, semenov1985} show a similar broadening effect.) \ The
lattice-based Flory theory allows for great versatility in describing a
variety of systems, including mixtures and various types of molecules. \
Flory and others used the lattice framework to extend models to semiflexible
rods \cite{semiflexible1956}, mixtures of rods and flexible coils \cite%
{flory1978_5}, rods connected by flexible linkers \cite{flory1978_6}, rods
with flexible side chains \cite{ballauff}, and, most relevant to the present
study, polydisperse rods \cite{flory1978_1}. \ However, the myriad numerical
approximations inherent in Flory's treatment have limited the quantitative
applications of the theory to anisotropic nanoparticles \cite{davisscience},
particularly those that interact differently at long and short ranges. \
Moreover, the \textquotedblleft attractive\textquotedblright\ interactions
in Flory's theory simply indicate the enthalpy of mixing. \ In a poor
solvent, the rods tend to \textquotedblleft attract\textquotedblright\ one
another simply because of the poor compatibility between rod and solvent. \
This attraction only acts on molecules immediately bordering the rod in the
lattice, i.e., at short ranges.

There have been an number of other efforts to describe attractive
interactions between rods. \ A number of studies have focused on depletion
forces caused by non-adsorbing polymers in solution with the rods. Surve 
\textit{et al.} modeled the effects of adsorption in a Flory-like framework 
\cite{surve}, while Bolhuis \textit{et al.} used particle-based simulations
to simulate spherocylinders with depletion forces caused by the addition of
nonadsorbing polymer \cite{bolhuis}. \ Borukhov \textit{et al.} studied
attractive depletion forces caused by cross-linkers in the context of stiff
biopolymers in aligned phases \cite{borukhov2001, borukhov2005}.

Particle-based simulations have also been used to simulate the phase
behavior of rigid rods, usually described as ellipsoids, but these
simulations are typically restricted to finite aspect ratio \cite{samborski}%
. \ The free energy of these systems are computed via Monte Carlo methods. \
Such simulations have been used to simulate the behavior of short aspect
ratio spherocylinders interacting through attractive temperature-dependent
square-well potentials \cite{lagoPRE, lagoJCP}. \ Because of the shorter
aspect ratio and the importance of temperature, these types of simulations
are often applied to alkanes \cite{williamson1999}, far smaller than
liquid-crystalline polymers or nanotubes; these simulations have also
detected the existence of a nematic-liquid-vapor triple point for these
shorter spherocylinders \cite{samborski1993, williamson1998}. \
Particle-based simulations provide a useful means of evaluating
approximations in Onsager-type frameworks; for instance, Samborski \textit{%
et al }used Monte\ Carlo simulations as a point of comparison for various
techniques of computing higher order virial coefficients for the Onsager
theory \cite{samborski}.

\subsection{Effects of Polydispersity}

In a system of rods, length polydispersity broadens the biphasic region, and
long rods preferentially enter the nematic phase, as intuited by Onsager. \
Moreover, the coexisting isotropic and nematic concentrations depend
strongly on the initial (or \textquotedblleft parent\textquotedblright )
concentration of rods. \ In a solution of monodisperse rods, these
coexistence concentrations are independent of the parent concentration; in a
solution of polydisperse rods, the increased degree of freedom in the
length-dimension yields much richer phase behavior. \ Initial studies of
polydispersity focused on long/short rod mixtures \cite{lekkerkerker}\ and
thick/thin rod mixtures \cite{shundyak}. \ Simple bidispersity in length or
diameter can have important effects, including the formation of
non-monotonic density profiles at the interface \cite{shundyak}.

Three papers in 2003 rigorously analyzed the effects of polydispersity for
continuous distributions of rigid rods. \ Speranza and Sollich computed the
coexistence concentrations of a system of Onsager rods with continuous
polydispersity in length \cite{speranza, speranzaJCP}, and they paid special
attention to the effects of the longest rods in the distribution. \ Wensink
and Vroege introduced an approximation for the excluded volume interaction
and systematically computed the biphasic boundaries as a function of the
initial concentration and the degree of polydispersity \cite{wensink}. \ \
Notably, they found that, as polydispersity increases, the isotropic
boundary moves to lower concentrations, and the nematic boundary shifts to
dramatically higher ones, broadening the biphasic region. \ These three
studies also identified specific scenarios where a triphasic (isotropic,
short rod nematic, long rod nematic) equilibrium state can occur. \ Although
the numerical analysis in \cite{wensink} was limited to two forms of the
parent population with varying polydispersity, the theory applies to systems
described by any rod length distribution.

These studies of polydisperse rod systems were limited to athermal solvents;
here we aim to extend this work to include the effect of long-range
attraction combined with short-range repulsion.

\subsection{Experimental Motivation}

The present study is motivated by recent experimental results on the phase
behavior of single-walled nanotubes\ (SWNTs)\ dispersed in superacids, such
as fuming sulfuric acid and chlorosulfonic acid \cite{Rai, Ramesh, Davis,
davisscience}. \ Since Iijima \textit{et al}'s observation of SWNTs in 1993 
\cite{IijimaSWNT}, SWNTs have attracted great interest because of their
unique electronic and mechanical properties \cite{baughmanscience2000}; much
current research is devoted to the creation of neat SWNT macroscopic
articles with properties comparable to the constituent SWNTs. \ However, the
production of such macroscopic articles has been hindered by the difficulty
of dispersing SWNTs as individual rods.

The dispersion of SWNTs is difficult because they tend to bundle tightly due
to strong van der Waals attraction to each other. \ Few solvents disperse
pristine SWNTs as individual rods capable of forming a concentrated
liquid-crystalline phase. \ SWNTs can form a liquid crystalline solution in
water when wrapped or stabilized by acids containing hydrophobic groups such
as DNA or hyaluronic acid \cite{badaire, moulton} or when trapped in
hydrogels \cite{nematicgel}. \ Superacids, such as fuming sulfuric acid and
chlorosulfonic acid, are the only fluids that spontaneously disperse SWNTs
as individual rods in a concentrated liquid crystalline state \cite{Rai,
Davis, davisscience}. \ Superacids protonate the sidewalls of SWNTs and
cause electrostatic repulsion and debundling \cite{Ramesh}. \ These liquid
crystalline dopes have been used for the production of macroscopic SWNT
articles \cite{Ericson, davisscience, behabtureview}.

Surprisingly, in superacids weaker than (roughly) equimolar mixtures of
sulfuric and chlorosulfonic acids, liquid-crystalline domains are in
equilibrium with a dilute isotropic phase. \ These liquid-crystalline
domains are long, threadlike structures,  termed \textquotedblleft SWNT
spaghetti,\textquotedblright\ composed by a myriad of aligned SWNTs spaced
by a few SWNT diameters. \ The SWNTs can translate along the axis of the
strand, but have very limited mobility in the normal direction. \ This
unusual phase behavior cannot be described by the Onsager or Flory theories
for monodisperse rods because the SWNTs are polydisperse and because the
theories cannot accurately describe the effects of varying acid strength on
the inter-rod interactions. \ In fact, the broadening of the biphasic
chimney occurs on the isotropic side rather than the nematic side, contrary
to Flory's predictions.

Modeling the phase behavior of such systems would involve both electrostatic
forces due to protonation and inter-SWNT van der Waals attraction. \ The
balance of electrostatic repulsive interactions and van der Waals attractive
interactions for dilute polyelectrolyte colloids have been modeled using the
Derjaguin, Landau, Verwey, and Overbeek (DLVO) theory \cite{dlvo, israelbook}%
. \ The DLVO approach was generalized for interactions between charged
anisotropic macromolecules by Chapot \textit{et al} \cite{chapot}. \ The
DLVO\ approach has been applied to a number of experiments on aqueous
suspensions of charged rods such as V$_{2}$O$_{5}$ gels \cite{pelletier} and
b-FeOOH \cite{maedamaeda}. DLVO and similar theories cannot be applied to
rods in superacids because the assumption of dilute electrolytes inherent in
the DLVO theory are inapplicable.

We postulate that the key physical mechanism controlling the behavior of
SWNTs in acids is the competition between long-range van der Waals
attraction, which is weakly dependent on acid strength, and short-range
electrostatic repulsion, which depends strongly on acid strength---as shown
clearly by Raman spectroscopy of SWNTs in acids. We approximate this
competition through a simple inter-rod repulsive-attractive potential, where
strength of the attractive well varies with solvent quality. \ The rods
interact through an attractive force that varies inversely with acid
strength; this attractive force mimics van der Waals interactions offset to
varying degrees by electrostatic repulsion as a function of protonation.

\section{Formulation}

\subsection{Free Energy}

Our formulation follows that of Wensink and Vroege and includes an
additional potential to capture the effect of attractive interparticle
interactions and solvent effects \cite{wensink}. \ A solution of thin rigid
rodlike molecules of diameter $d$ and polydisperse lengths $L$ is described
by a rod distribution function $N(l)$. $N(l)$ refers to the number of rods
in the system with relative length $l$, where $l$ is defined as $L/L_{0}$
and $L_{0}$ is an arbitrary reference length. \ For a volume $V$, $N(l)/V$
is the local number density of rods of length $l$, and $\int N(l)dl$ gives
the total number of rods in the system.

The dimensionless Onsager excess free energy $f$ for a volume $V$ is written
as 
\begin{equation*}
f=\frac{bF}{Vk_{B}T}\sim \dint c(l)[\ln c(l)-1]dl+\dint c(l)\omega (l)dl
\end{equation*}%
\begin{equation}
+\dint \dint c(l)c(l^{\prime })ll^{\prime }[\rho (l,l^{\prime })+\lambda
(l,l^{\prime })]dldl^{\prime },
\end{equation}%
where the reference volume $b$ is defined as $\pi dL_{0}^{2}/4$, and the
dimensionless rod distribution function $c(l)$ is defined as $bN(l)/V$. \
The total rod number concentration is $c_{0}$, the zeroth moment of the
distribution $c(l)$, where distribution moments are defined as 
\begin{equation}
c_{k}=\int c(l)l^{k}dl\mathbf{.}
\end{equation}%
The overall concentration of the system can also be described by the rod
volume fraction $\phi =(d/L_{0})c_{1}$. \ The rod volume fraction is a more
experimentally accessible quantity.

The entropic terms in the free energy description can be described as
follows. \ The first term describes the ideal free energy of the
polydisperse system. \ The second term contains the orientational entropy
contribution,%
\begin{equation}
\omega (l)=\int \psi (l,\mathbf{u)}\ln [4\pi \psi (l,\mathbf{u})]d\mathbf{u,}
\end{equation}%
where $\psi (l,\mathbf{u)}$ is the normalized angular distribution function
for rods with relative length $l$ with orientation described by the unit
vector $\mathbf{u}$.\ \ This term is minimized by a uniform (isotropic)
distribution of rod angles written as $\psi =1/4\pi $. \ The third term
contains the excluded volume interactions between rods. \ The average
excluded volume contribution from rods of relative length $l$ and $l^{\prime
}$ is 
\begin{equation}
\rho (l,l^{\prime })=\frac{4}{\pi }\int |\mathbf{u\times u}^{\prime }|\psi
(l,\mathbf{u)}\psi (l^{\prime },\mathbf{u}^{\prime })]d\mathbf{u}^{\prime }d%
\mathbf{u.}
\end{equation}%
This term equals 1 for an isotropic state and is minimized by an aligned
nematic state where the rods' alignment minimizes their excluded-volume
interactions. \ The attractive/repulsive potential $\lambda (l,l^{\prime })$
models inter-rod attraction and repulsion, and is new. \ The phase behavior
of idealized Onsager rods is governed by the balance between $\omega (l)$
and $\rho (l,l^{\prime })$ as a function of $c(l)$.

We follow Wensink and\ Vroege in using the uniaxial Gaussian ansatz to
approximate the angular distribution function $\psi (l,\mathbf{u)}$, written
as%
\begin{equation}
\psi (l,\mathbf{u)}=\left\{ 
\begin{array}{ccc}
\frac{\alpha (l)}{4\pi }\exp \left[ -\frac{1}{2}\alpha (l)\theta ^{2}\right] 
& \text{for} & 0\leq \theta \leq \pi /2 \\ 
\frac{\alpha (l)}{4\pi }\exp \left[ -\frac{1}{2}\alpha (l)(\pi -\theta )^{2}%
\right]  & \text{for} & \pi /2\leq \theta \leq \pi 
\end{array}%
,\right.   \label{kappa}
\end{equation}%
such that $\psi (l,\mathbf{u)}$ is reduced to an unknown in only the $l$%
-dimension through the function $\alpha (l)$. \ The function $\alpha (l)$ is
left as an unknown that describes the degree of alignment in a given phase.
\ Because of their higher excluded-volume cost, longer rods tend to be more
aligned. \ Wensink and Vroege write $\omega (l)\sim \ln \alpha (l)-1$ and
then develop an asymptotic expansion of $\rho (l,l^{\prime })$ for the
nematic phase in terms of $\alpha $ as%
\begin{equation}
\rho (l,l^{\prime })\sim \sqrt{\frac{8}{\pi }\left( \alpha (l)^{-1}+\alpha
(l^{\prime })^{-1}\right) }\mathbf{,}
\end{equation}%
where terms involving higher powers of $\alpha (l)^{-1}$ are neglected.

We designate variables as pertaining to the isotropic phase, nematic phase,
and parent phase by using the superscripts (I), (N), and (0) for the
theoretical description of a parent population $c^{(0)}(l)$\ phase
separating into $c^{(I)}(l)$ and $c^{(N)}(l)$.

\subsection{Equilibrium Conditions}

The degree of alignment for any nematic equilibrium state must minimize the
free energy. \ In the present case, the degree of alignment is expressed
through the unknown $\alpha (l)$, and the condition is written as%
\begin{equation*}
\frac{\delta f}{\delta \alpha (l)}=0=\frac{c(l)}{\alpha (l)}-\left( \frac{8}{%
\pi }\right) ^{1/2}\frac{lc(l)}{\alpha ^{2}(l)}\int l^{\prime }c(l^{\prime
})\left( \alpha (l)^{-1}+\alpha (l^{\prime })^{-1}\right) ^{-1/2}dl^{\prime }
\end{equation*}%
\begin{equation}
+lc(l)\int l^{\prime }c(l^{\prime })\frac{\delta \lambda (l,l^{\prime })}{%
\delta \alpha (l)}dl^{\prime }\mathbf{.}  \label{alpha}
\end{equation}%
The excluded volume term in Wensink and Vroege's equation 7 is incorrect by
a factor of two, while their equation 8 is correct. \ We argue below that
the function $\lambda (l,l^{\prime })$ is a weak function of $\alpha (l)$
for the nematic phase such that the final term may be neglected. \ For an
isotropic-nematic coexistence state to exist at equilibrium, two additional
criteria must be satisfied. \ The two phases must have equal chemical
potentials [$\mu (l)/k_{B}T=\delta f/\delta c(l)$] as a function of $l$,
yielding the condition 
\begin{equation}
\ln c^{(I)}(l)+2lc_{1}^{(I)}=\ln c^{(N)}(l)+\ln \left[ \alpha (l)\right]
-1+\mu _{ex}^{(N)}(l)\mathbf{,}  \label{chemicalpotential}
\end{equation}%
and the excess chemical potential for the nematic phase $\mu _{ex}^{(N)}(l)$
is defined as%
\begin{equation}
\mu _{ex}^{(N)}(l)=2l\int l^{\prime }c^{(N)}(l^{\prime })\lambda
(l,l^{\prime })dl^{\prime }+2l\int c^{(N)}(l^{\prime })l^{\prime }\rho
(l,l^{\prime })dl^{\prime },
\end{equation}%
\ This criteria must be fulfilled for all $l$. \ The two phases must also
have the same osmotic pressure [$\Pi =\delta f/\delta V$] which yields the
additional equation%
\begin{equation}
c_{0}^{(I)}+\left( c_{1}^{(I)}\right) ^{2}=c_{0}^{(N)}+\int
lc^{(N)}(l)l^{\prime }c^{(N)}(l^{\prime })[\rho (l,l^{\prime })+\lambda
(l,l^{\prime })]dldl^{\prime }.  \label{osmotic}
\end{equation}

Finally, conservation of mass must hold such that the two daughter phases in
coexistence must add together to the parent population, i.e.,%
\begin{equation}
c^{(0)}(l)=(1-\gamma )c^{(I)}(l)+\gamma c^{(N)}(l),  \label{mass}
\end{equation}%
where $\gamma $ is the volumetric fraction of the system that is nematic. \
Given a starting population $c^{(0)}(l)$ and attractive potential $\lambda
(l,l^{\prime })$, Eqs. (\ref{alpha}, \ref{chemicalpotential}, \ref{osmotic}, %
\ref{mass}) comprise three functional equations and one scalar equation that
may be solved for $c^{(I)}(l)$, $c^{(N)}(l)$, $\alpha (l)$, and $\gamma $. \
(Alternatively, the normalized parent population $c^{(0)}(l)/c_{0}^{(0)}$\
and $\gamma $ may be specified with the quantity $c_{0}^{(0)}$\ as an
unknown.) \ For convenience, we take the constant reference length $L_{0}$\
to be $L_{mean}$, the average length of the parent population.

The Parsons scaling approach allows for use of the Onsager excluded volume
at high concentrations \cite{semenov1985}. \ The difference in excluded
volume is less than 5\% for $\phi <0.2$. \ We adapt the Parsons scaling
approach for polydisperse systems to write the excluded volume term of the
free energy as 
\begin{equation}
\frac{bF_{Parson}}{Vk_{B}T}\sim \int c(l)\left( \frac{L_{0}}{d}\right)
l(-\ln (1-\phi (l^{\prime }))\rho (l,l^{\prime })dldl^{\prime }.
\end{equation}%
After some mathematical manipulation, the contributions to the chemical
potential and osmotic pressure are then written as%
\begin{equation}
\Pi _{Parson}\sim \int \frac{c(l)c(l^{\prime })ll^{\prime }\rho (l,l^{\prime
})}{(1-\phi (l^{\prime }))}dldl^{\prime }.
\end{equation}%
\begin{equation}
\mu _{Parson}\sim \left( \frac{L_{0}}{d}\right) l\int (-\ln (1-\phi (l))\rho
(l,l^{\prime })dl^{\prime }+\frac{l}{(1-\phi (l^{\prime }))}\int c(l^{\prime
})l^{\prime }\rho (l,l^{\prime })dl^{\prime }.
\end{equation}%
The equation for equilibrium alignment becomes%
\begin{equation*}
\frac{\delta f}{\delta \alpha (l)}=0=\frac{2c(l)}{\alpha (l)}-\left( \frac{8%
}{\pi }\right) ^{1/2}\left( \frac{L_{0}}{d}\right) \frac{lc(l)}{2\alpha
^{2}(l)}\int (-\ln (1-\phi (l^{\prime }))\left[ \left( \alpha
(l)^{-1}+\alpha (l^{\prime })^{-1}\right) ^{-1/2}+\frac{\delta \lambda
(l,l^{\prime })}{\delta \alpha (l)}\right] dl^{\prime }
\end{equation*}%
\begin{equation}
-\left( \frac{8}{\pi }\right) ^{1/2}\left( \frac{L_{0}}{d}\right) \frac{%
(-\ln (1-\phi (l))}{2\alpha ^{2}(l)}\int l^{\prime }c(l^{\prime })\left[
\left( \alpha (l)^{-1}+\alpha (l^{\prime })^{-1}\right) ^{-1/2}+\frac{\delta
\lambda (l,l^{\prime })}{\delta \alpha (l)}\right] dl^{\prime },
\end{equation}

\subsection{Attractive and Repulsive Interactions}

The novel contribution to the free energy expression is expressed in the
quantity $\lambda (l,l^{\prime })$. \ We propose a particular
phenomenological model inspired by the unusual \textquotedblleft SWNT
spaghetti\textquotedblright\ liquid crystalline phase illustrated in Figure
1; the SWNTs in these strands are aligned along the strand with small
distances between SWNT sidewalls. \ The SWNTs are no longer in due to the
short-range electrostatic repulsion, but the attractive forces keep them
within a few diameters of each another, such that each SWNT interacts with
an effective \textquotedblleft cage\textquotedblright\ of other SWNTs within
the liquid crystalline phase.

Our proposed model is as follows: \ A rod in an aligned state is subject to
an attractive interaction with surrounding rods $u(D)$, which represents the
energy per unit length of the rod as a function of the distance $D$ between
the centers of the aligned rods. \ This quantity scales linearly with $l$
and is made dimensionless as $U(D)\equiv u(D)L_{mean}/k_{B}T$ . \ The
square-well potential $U(D)$ is depicted in Figure 1 and is characterized by
the four parameters $U_{\max }$, $U_{\min }$, $D_{1}$, and $D_{2}$. \ The
discontinuities in the derivatives of the square-well potential are made
continuous by using an arctan function to approximate the transitions of $%
U(D)$ across $D_{1}$ and $D_{2}$.

As a first approximation, this attractive interaction is applied only to the
dense nematic phase. \ This is motivated by experimental observations of the
unusual liquid-crystalline phases in SWNT/superacid systems. The liquid
crystalline phase is marked by aligned SWNTs in long threadlike domains with
acid molecules in-between, and the inter-SWNT distances are on the order of $%
10-30nm$. This means that rods in the nematic phase are separated by average
distances that are much smaller than the average distances between rods in
the isotropic phase. This means that the rods in the nematic phase will
spend substantially more time in the attractive well than rods in the
isotropic phase; we approximate these experimental observations simply by
applying the attractive well to the nematic phase but not the isotropic
phase. The attractive interaction varies weakly with alignment in nematic
phases. \ On average, the nematic phase will behave much like a columnar
phase where the average inter-rod distance $D$ can be calculated from $\phi
^{(N)}$, the rod volume fraction in the nematic phase, as 
\begin{equation}
D=\frac{d}{2}\left( \frac{\phi ^{(N)}\sqrt{3}}{2\pi }\right) ^{-1/2}\mathbf{.%
}
\end{equation}%
As the \textquotedblleft test\textquotedblright\ rod feels the attractive
potential, the rod's nearest neighbors in the densely packed nematic phase
will \textquotedblleft shield\textquotedblright\ the rod from others that
are farther out such that the attractive interaction is minimal beyond those
nearest neighbors. \ (Also, those nearest neighbors will be the first to
enter the repulsive region.) \ Consequently, the quantity $lc(l)\int
l^{\prime }c(l^{\prime })\lambda (l,l^{\prime })dl^{\prime }$ is rewritten
as $lc(l)\Lambda (l)$ where $\Lambda (l)=6U(D)$.\textbf{\ }\ (A similar
argument is used by Wensink and Vroege in the case of average excluded
volume in the nematic phase.) \ The square-well potential can be
parameterized by the dimensionless quantities $U_{\max }$, $U_{\min }$, $%
D_{1}/d$, and $D_{2}/d$, which may vary with solvent quality. \ In addition,
the repulsive region ($D<D_{1}$) increases the effective excluded volume by
a factor of $(D_{1}/d+1)/2$. \ The repulsive portion of the square-well
potential is applied to both isotropic and nematic phases.\textbf{\ }

\subsection{Shadow and Cloud Phases}

The phase behavior of polydisperse rod populations is complicated by the
fact that the parent population $c^{(0)}(l)$ influences the properties of
the daughter populations $c^{(I)}(l)$ and $c^{(N)}(l)$. \ This stands in
contrast to the monodisperse case where the coexistence values of $%
c_{0}^{(I)}$ and $c_{0}^{(N)}$ are independent of the parent concentration $%
c_{0}^{(0)}$. \ These concepts are illustrated in Fig. 2, which depicts the
dependence of $c_{0}^{(I)}$ and $c_{0}^{(N)}$ on $c_{0}^{(0)}$. A thorough
explanation of cloud and shadow phases may be found in the appendix.  \ 

\subsection{Parent Distribution and Solution Method}

For the SWNT samples of interest, the parent rod population is
well-described by a simple Schultz distribution 
\begin{equation}
c^{(0)}(l)/c_{0}^{(0)}=Nl^{z}\exp [-(z+1)l],
\end{equation}%
with normalization factor $N$ and is truncated at some minimum and maximum $%
l $. \ The polydispersity of any population is described by the parameter $%
\sigma $, defined as 
\begin{equation}
\sigma ^{2}=c_{2}^{(0)}/\left( c_{1}^{(0)}\right) ^{2}-1,
\end{equation}%
which is related to the parameter $z$ in the Schultz distribution as $\sigma
^{2}=(1+z)^{-1}$.

In a typical purified batch of HiPco SWNTs (HPR batch 152.2), we find that
the parent distribution of SWNTs is well-approximated through the Schultz
distribution as follows: \ $d=1nm$, $L_{mean}=348nm$, $L_{\min }=17nm$ (the
SWNT purification process eliminates SWNTs below a particular cutoff), $%
L_{\max }=8\mu m$, and $\sigma =0.6$ \cite{davisthesis}. \ \ We use this as
an example parent population.

In general, we explore numerically the phase behavior of a SWNT-dispersion
as a function of the attractive potential well depth $U_{\min }$,
concentration $\phi _{0}$, and of the polydispersity $\sigma $. \ For all
studies, we set $U_{\max }\gg |U_{\min }|$ since variations in $U_{\max }$
do not substantially change the phase behavior as long as $U_{\max }>1$, and
we set $D_{2}/d\gg 1$ because nematic phases where $D>D_{2}$ do not differ
from the athermal case.

The $l$-space is discretized into $Q$ points in a fashion similar to that
described by Wensink and Vroege, and Eqs. (\ref{alpha}, \ref%
{chemicalpotential}, \ref{osmotic}, \ref{mass}) are discretized into $3Q+1$
nonlinear equations and solved simultaneously using Newton's method. \
Solutions are traced in $c_{0}^{(0)}$-space (or $\gamma $-space), $\sigma $%
-space and $U_{\min }$-space using arclength continuation in the parameter
of interest \cite{bolhuis}. \ The following results and discussion are
specific to the Schultz distribution but should be representative of the
behavior of other parent populations. \ (A typical value for $Q$\ is $70$.
Note that the points need not be evenly spaced.) 

\section{Results \&\ Discussion}

\subsection{Cloud curves}

For the example population of rods described above with $\sigma =0.6$, the
isotropic and nematic cloud curves are computed as a function of $U_{\min }$
for the case where $D_{1}/d=1$ (Fig. 3). \ $U_{\min }=0$ corresponds to
polydisperse athermal rods; the isotropic cloud point is $\phi ^{(I)}=0.0049$%
, and the nematic cloud point is $\phi ^{(N)}=0.177$.

As attractive interactions grow, the nematic phase becomes more
thermodynamically favorable and more concentrated. Because $D_{1}/d=1$, the
aligned rods can pack tightly together in the liquid crystalline phase, so
the nematic cloud curve moves rapidly to the maximum packing fraction as $%
U_{\min }$ decreases. \ Setting $D_{1}/d>1$ essentially sets a minimum for
inter-rod spacing and a maximum $\phi ^{(N)}$.\ Such a case is shown as a
dotted line\ in Fig. 3 for $D_{1}/d=2.08$, where the increase in $\phi ^{(N)}
$ is arrested by short-range repulsion.

The nematic cloud curve in Fig. 3 \ for $D_{1}/d=1$ is similar to the
classic Flory theory \cite{flory1956} and the Khokhlov extension of the
Onsager theory \cite{LCinpolymers} because the biphasic region primarily
broadens by increasing the concentration of the nematic cloud curve $\phi
^{(N)}$ toward the maximum packing fraction (0.907) as the solvent quality
decreases. \ In all these previous studies, the nematic concentration
rapidly increases to a solid-like state because short-range repulsion is not
included in the model. \ The Parsons scaling approach causes the cloud curve
to level off, similar to the results of Khokhlov\textit{\ et al}.

Most interestingly, the isotropic cloud curve diverges to very small values
at a critical value of $U_{\min }^{\ast }\cong -0.581$. \ Because $\log
c^{(I)}(l)\sim l\phi ^{(N)}U_{\min }$, the quantity $\log \phi ^{(I)}$\ will
become more and more negative as the attractive force in the nematic phase
increases.\ Therefore in systems where attractive interactions are stronger
than this critical value, the dilute, isotropic phase must always coexist
with some liquid-crystalline phase. \ Even at very small concentrations, the
system is still in the biphasic regime; thus, some rods will form an aligned
phase at equilibrium due to the strong attraction. \ A similar result was
found by Speranza and Sollich; for distributions with small amounts of
extremely long rods, the cloud curve decreased to vanishingly small
concentrations \cite{speranza}.  \ This holds great import for experimental
studies of rod dispersions, as discussed below.

Figures 4 and 5 show the influence of polydispersity $\sigma $ on the
isotropic and nematic cloud curves for $D_{1}/d=1$.

As polydispersity decreases, the isotropic cloud curves move down to lower
well depths and to the right to higher rod concentration; this is to be
expected since the effects of length fractionation decrease as
polydispersity decreases. \ This also means that $U_{\min }^{\ast }$
decreases as a function of $\sigma $; this holds practical import for
experimental scenarios where an isotropic solution is desired. \ If a
particular solvent/rod combination corresponds to a value of $U_{\min }$
that is less than $U_{\min }^{\ast }$, then even dilute solutions will be in
the biphasic regime. \ If the polydispersity of the solution can be
decreased and/or the average length increased, then the value of $U_{\min
}^{\ast }$ will also decrease; thus, polydispersity can be used as a control
to ensure that $U_{\min }>U_{\min }^{\ast }$.

As polydispersity decreases, the nematic cloud curve moves down to lower
well depths and to the left to lower rod concentration. \ For low $\sigma $,
the nematic cloud curve is a much weaker function of acid strength than the
isotropic cloud curve. \ Thus, for broad distributions (e.g., $\sigma =0.6$%
), the nematic cloud curve varies more rapidly than the isotropic cloud
curve while for narrow distributions (e.g., $\sigma =0.4$) the opposite is
true. \ Note that in the monodisperse case, the nematic cloud curve actually
decreases with the attractive well becomes stronger; this differs from the
polydisperse case because the attractive well amplifies the effects of
polydispersity and broadens the biphasic region. 

\subsection{Comparison with experiment}

In previous papers, we used a series of experiments to establish a phase
diagram for SWNT/acid mixtures \cite{Rai, davisscience} which is displayed
in Figure 6. \ Various SWNT/acid mixtures at a given SWNT concentration
(denoted by diamonds in Figure 6) are centrifuged into isotropic and nematic
phases, and the concentration $\phi ^{(I)}$ of the isotropic phase is
measured\footnote{%
Concentration is measured using UV-vis-NIR absorbance.\cite{Rai}} and marked
as black circles.\ The concentration $\phi ^{(I)}$ grows with increasing
fractional charge (better solvent quality). \ In contrast with the
predictions of Flory-like theories, the dilute isotropic phase in poor
solvents coexists with a concentrated liquid-crystalline phase ($\phi
^{(N)}\approx 0.1$)\ rather than with a solid phase (i.e., $\phi ^{(N)}>0.9$%
).

In order to compare the theoretical results with experimental results for
SWNTs in superacids, a relationship between $U_{\min }$\ and acid strength
must be posited. \ Superacid strength is typically measured in terms of\
base-specific Hammett acidity \cite{pross}. \ In the case of SWNT/superacid
systems, acid strength is related to the fractional charge per carbon caused
by protonation of the nanotube sidewall \cite{Rai}. \ (This quantity varies
linearly with the wavenumber shift, $dG$, of the G-peak in the SWNT Raman
spectra.) \ The van der Waals forces should be roughly independent of acid
strength. \ However, as the degree of protonation increases, the repulsion
effect increases. \ Thus, $U_{\min }$, which represents the sum of the van
der Waals forces and the repulsive forces, becomes shallower as acid
strength increases.\footnote{%
A similar effect is observed in DLVO theory for a fixed attractive force but
a repulsive force that increases with increasing solvent quality.} \ We
approximate the dependence of the repulsive force on the acid strength by
positing a simple linear relationship between $dG$\ and the attractive well
depth $U_{\min }$\ by comparing the theoretical and experimental results for
the isotropic concentration after phase separation.\footnote{%
A number of factors will influence the relationship between acid strength
and $U_{\min }$, and these factors may vary from one SWNT sample to another.
\ $U_{\min }$ is independent of SWNT length and polydispersity, but
experiments indicate that the solubility of a HiPco SWNT sample in a given
acid varies with diameter distribution, the frequency of defects in the SWNT
sidewalls, and the chemical effects of the oxidation and purification
process on the SWNT sidewalls.}

Experimental measurements for the strongest acid used (chlorosulfonic acid)
indicate that $\phi ^{(I)}$ is approximately equal to the theoretical
predictions for athermal (Onsager) rods; thus, we match a Raman shift of $%
dG=25cm^{-1}$ to $U_{\min }=0$. \ The theoretical predictions for $\phi
^{(I)}$ for $U_{\min }=-0.957$ match the values of $\phi ^{(I)}$ from
centrifugation experiments on SWNTs in 102\% H$_{2}$SO$_{4}$ (i.e., 100\% H$%
_{2}$SO$_{4}$ with excess 2 wt\% dissolved SO$_{3}$) which has a Raman shift
of $dG=17cm^{-1}$. \ These two data points set the linear relationship
between $U_{\min }$ and $dG$.

Figure 6 combines computational and experimental results in a full phase
diagram for SWNT/acid mixtures. \ The model (red open circles) is remarkably
successful at predicting the experimental values $\phi ^{(I)}$ (black
circles) at low concentrations \cite{davisscience}. \ Therefore, the simple
square-well potential captures the chief aspects of the physical
interactions between SWNTs in acids. \ 

Additional experiments were performed using the methods of Rai \textit{et al}
to test the model \cite{Rai}. \ The concentration $\phi ^{(I)}$ of the
isotropic phase in 120\%\ H$_{2}$SO$_{4}$ was measured as a function of the
starting concentration $\phi ^{(0)}$. \ A comparison between theoretical and
experimental data points is displayed in Figure 7. \ Again, the agreement
between the two shows that the fundamental physics of inter-nanotube
interactions are captured by the theory.

Experimental measurements indicate a nematic cloud point of $\phi
^{(N)}\approx 0.1$ that is relatively constant with respect to acid
strength; however, the theoretical results indicate a slightly constant
value of $\phi ^{(N)}=0.115$. \ Unlike the case of low-concentration phase
separation, the theoretical and experimental values for the nematic cloud
curve do not match. \ However, further analysis of the theoretical results
reveals why the values differ. \ In Fig. 8, the theory predicts that $\gamma
\approx 0.99$ for a system where $\phi ^{(0)}=0.1$, i.e., $99\%$ of the
system (by volume)\ is nematic. \ The remaining $1\%$ is isotropic and
comprised of relatively short rods, as shown in Fig. 8. \ In an experiment,
these short-rod-dominated isotropic regions would be found primarily in the
defects between liquid-crystalline domains, and these small isotropic
regions would be nearly undetectable by optical microscopy, rheology, or
dynamic scanning calorimetry. \ Thus, the experimental points are best taken
as measurements of the transition to a system that is $99\%$ liquid
crystalline by volume, and we may take the theoretical prediction of the
nematic cloud point as correct. \ The theory predicts that the starting
concentration must be pushed up to $\phi ^{(0)}=0.115$ to completely
eliminate these isotropic regions. \ However, simulation results also
indicate that if all of the rods with an aspect ratio under $40$ are
removed, then the nematic cloud curve decreases below  $\phi ^{(0)}\approx
0.1$.\footnote{%
We do not believe there is nematic-nematic separation in our experimental
system. Wensink and Vroege indicate that for lognormal distributions, there
is a fairly small window of $\sigma $\ and $\phi ^{(0)}$\ where this is
possible.}

\subsection{Applications}

The computational results show that a liquid-crystalline phase will always
be present for acids with $dG\,<20.15cm^{-1}$ ($U_{\min }<U_{\min }^{\ast }$%
, where $U_{\min }^{\ast }=-0.581$, equivalent to an acid mixture with a
9:16 volume ratio of ClSO$_{3}$H to H$_{2}$SO$_{4}$), regardless of the
starting SWNT concentration. \ This means that in order to get a dilute,
isotropic suspension of individual SWNTs in superacid, the solution must
first be centrifuged and phase separated in order to remove the isotropic
phase from the nematic phase. \ A number of CNT characterization techniques
(such as rheological characterization of CNT length \cite{nickpv}) require a
dilute, isotropic suspension, and these counterintuitive computational
results show that a simple low-concentration dispersion of SWNTs in 120\% H$%
_{2}$SO$_{4}$ will not yield a dilute, isotropic solution. \ The isotropic
cloud curve can be moved down to lower well depths and to the right to
higher rod concentrations by eliminating the longest SWNTs in a given
sample; this would be useful for applications requiring a dilute, isotropic
solution and could be accomplished via a number of experimental techniques.

Also, one of the chief applications of SWNT / superacid dispersions is the
formation of aligned articles such as fibers and films \cite%
{Ericson,davisscience}; our computational results have immediate
applications for the processing of such articles. \ The fiber spinning
process works as follows:\ \ High-concentration, liquid-crystalline SWNT /
superacid dispersions are mixed, and extruded. \ The viscous dispersion
becomes more aligned due to the tension on the fluid during extensional
flow. \ Finally, the dispersion is coagulated in a non-solvent bath where
the acid is removed, and the SWNT dispersion solidifies to form an aligned
fiber. \ The best fibers are produced by dispersions that are highly aligned
and fully liquid-crystalline, i.e., above the nematic cloud curve. \ Any
isotropic regions in the dispersions form misaligned, weak regions in the
final fiber; these defects decrease the electrical properties and are the
primary points of fiber mechanical failure. \ (Note that defects that are
predominantly composed of small nanotubes have already been observed for
liquid-crystalline MWNT systems \cite{ZhangKinloch}.)

With this application in mind, the theoretical framework outlined above can
be used to locate the true nematic cloud curve in order to ensure that the
dispersion is fully liquid crystalline. \ In this case, the cloud curve
concentration may be too high for practical mixing and fiber spinning, so
the theoretical framework can also be used to test the results of
manipulating the starting SWNT length distribution in order to decrease the
nematic cloud concentration. \ In this case, the elimination of SWNTs under $%
40$nm will decrease the nematic cloud concentration below $\phi
^{(0)}\approx 0.1$ such that SWNT/superacid dopes of $\phi ^{(0)}\approx 0.1$%
\ will be fully liquid crystalline.\footnote{%
This may be accomplished by a numerical of experimental techniques. \ These
include washing the initial SWNT sample with a weaker acid (such as 98\% H$%
_{2}$SO$_{4}$) to eliminate the shortest SWNTs. \ Also, fractionating a
sample in the biphasic regime in a weaker superacid (such as 102\% H$_{2}$SO$%
_{4}$) and removing the isotropic phase can be used to eliminate the
shortest SWNTs.} \ \ 

Also, this theoretical framework has implications for high-concentration
rheology of SWNT dispersions. \ The dependence of viscosity on concentration
is a well-known experimental test for liquid-crystallinity; the viscosity
increases, goes through a maximum, decreases due to the formation of a
liquid-crystalline phase, and finally begins to increase again at high
concentrations \cite{Davis}. \ This dependence on concentration is blurred
and weakened as the biphasic chimney is broadened because the formation of
aligned phase occurs over a much wider concentration range. \ SWNT/superacid
dispersions with weaker acids or polydisperse length distributions will show
a weak dependence of viscosity on concentration and may not show the
maximum/minimum signature of liquid-crystallinity. \ Decreasing
polydispersity or strengthening the acid solvent will sharpen and amplify
the difference between the maximum and minimum in viscosity.

The utility of the theoretical treatment outlined in this study is evident
from the unexpected nature of the results. \ Optical microscopy experiments
and DSC could not detect the isotropic regions in the SWNT/superacid
dispersions used for fiber spinning at $\phi ^{(0)}\approx 0.1$, but the
theory predicts that these regions are indeed present; such regions have
adverse effects on as-spun fibers. \ Also, it is entirely intuitive to
reason that a dilute solution of SWNTs in 120\% H$_{2}$SO$_{4}$ would be
dispersed as isotropic individuals, but the theory predicts that even these
low concentrations are within the biphasic region because of attractive
interactions.

We expect that this analysis will pave the way for the modeling of other
anisotropic materials such as multi-walled carbon nanotubes (MWNTs) or
inorganic nanorods. \ The phase behavior of CdSe nanorods \cite%
{cdse-phasediagram} has been assessed through purely qualitative means, and
an effective theoretical framework for understanding these materials is
needed. \ The phase transitions for solutions of polydisperse functionalized
MWNTs were qualitatively characterized by Song and Windle \cite{song,
song-kinloch}. \ A theoretical analysis of these systems may allow a more
quantitative understanding of these dispersed nanomaterials.

\subsection{Conclusions}

We have developed an extension to the Onsager theory for polydisperse rigid
rods developed by Wensink and Vroege in order to capture the balance of
long-range attractive and repulsive forces observed for anisotropic
nanomaterials in solution. \ Our work is particularly motivated by the
recent quantitative experimental data for SWNTs dispersed as individual
rigid rods in superacids. \ Our results indicate excellent agreement between
the theory and experimental data for predicting phase separation at low
concentrations and for predicting the biphasic chimney's broadening on the
isotropic side. \ The theoretical results also hold important and surprising
implications for a variety of SWNT/superacid experiments and applications,
including the understanding of SWNT/superacid rheology and the processing of
fibers and films from liquid crystalline dopes.

\textbf{Acknowledgments}

We acknowledge the help of Virginia Davis, Pradeep Rai, Valentin Prieto,
Howard Schmidt, Robert Hauge, Rick Smalley, and Wade Adams. \ Funding was
provided by the Office of Naval Research under Grant N00014-01-1-0789, AFOSR
grant FA9550-06-1-0207, AFRL agreements FA8650-07-2-5061 and
07-S568-0042-01-C1, NSF CAREER, and the Evans-Attwell Welch Postdoctoral
Fellowship. This material is based on research sponsored by Air Force
Research Laboratory under agreement number FA8650-07-2-5061.

\textbf{Appendix}

Two\ useful but oft-misunderstood concepts for delineating the phase
behavior of polydisperse rod populations are the cloud phase and shadow
phase. \ For example, at small concentrations, a parent population of
polydisperse rods in an athermal solvent will remain isotropic with no phase
separation. \ As the parent concentration $c_{0}^{(0)}$ is increased, a
point is reached where an infinitesimal volume of nematic phase, termed a
\textquotedblleft shadow phase,\textquotedblright\ forms while the vast
majority of the parent population remains in the \textquotedblleft
cloud\textquotedblright\ isotropic phase. \ Thus, the shadow nematic phase
coexists with the cloud isotropic phase, and the corresponding value of $%
c_{0}^{(0)}$\ represents the lowest value of $c_{0}^{(0)}$\ at which
coexistence is possible. \ The converse is true on the high-concentration
side of the phase diagram; the shadow isotropic phase coexists with the
cloud nematic phase, and the corresponding value of $c_{0}^{(0)}$\
represents the highest value of $c_{0}^{(0)}$\ at which coexistence is
possible. \ On the isotropic cloud curve, the quantity $\gamma $ (the
fraction of the system that is nematic) will approach zero, while on the
nematic cloud curve, $\gamma $ will approach one.

The isotropic and nematic cloud phases are the ultimate boundaries of the
biphasic region in a system of polydisperse rods. \ Parent populations
within the biphasic region will phase separate into coexisting isotropic and
nematic phases, and the values of $c_{0}^{(I)}$ and $c_{0}^{(N)}$ will vary
with $c_{0}^{(0)}$. \ These concepts are illustrated in Fig. \ref%
{illustration}, which depicts the dependence of $c_{0}^{(I)}$ and $%
c_{0}^{(N)}$ on $c_{0}^{(0)}$. \ Note that the daughter phases appear to
remain in the biphasic regime which suggests that they are not an
equilibrium solution; this paradox is resolved by the fact that the
daughters have a different length distribution than the parent phase, such
that the daughter phases lie outside the cloud curves for their specific
distribution.

\textbf{Figure captions}

Figure 1: \ (I) The combination of short range electrostatic repulsive
forces (shown in blue) and van der Waals attraction will result in an
attractive well (shown in red) potential between the two particles. However,
the exact form of these functions is unknown for complex solvent and solutes
such as superacids and SWNTs. \ (II) A simple, phenomenological square well
potential $U(D)$\ is used to capture this balance of repulsive forces and
attractive forces between SWNTs in various acids. \ As the acid quality
decreases, the well becomes deeper. \ (III) $D$\ is the distance between
neighboring SWNTs in the liquid-crystalline phase. This simplified potential
is inspired by the unusual liquid-crystalline order seen in
\textquotedblleft spaghetti\textquotedblright\ in SWNT/superacid systems 
\cite{Davis}. \ (IV) A schematic of \textquotedblleft
spaghetti\textquotedblright\ geometry depicts the threadlike, aligned nature
of these unusual liquid-crystalline domains.

Figure 2: \ Computational results for an athermal solvent illustrating the
dependence of $\phi ^{(I)}$\ and $\phi ^{(N)}$\ on $\phi ^{(0)}$\ for
polydisperse distributions for or $\sigma =0.6$, $L_{mean}=348d$. \ Squares
indicate shadow points, and diamonds indicate cloud points. \ This stands in
stark contrast to the monodisperse case where $\phi ^{(I)}$\ and $\phi ^{(N)}
$\ do not vary.

Figure 3: \ The isotropic and nematic cloud points are depicted as a
function of the well depth $U_{\min }$\ \ for $\sigma =0.6$, $L_{mean}=348d$%
.\ When $D_{1}/d=1$, the nematic cloud point rapidly increases to $\phi
^{(N)}=1$\ as $U_{\min }$\ decreases. \ For the case $D_{1}/d=2.08$, $\phi
^{(N)}$\ remains constant at $0.210$. \ The Onsager predictions for
monodisperse rods are depicted as triangles ($\bigtriangleup $).

Figure 4: \ Isotropic cloud curves ($U_{\min }$\ vs. $\phi $) as a function
of $\sigma $\ for $D_{1}/d=1$. \ As $\sigma $\ decreases, the isotropic
cloud curves move down to lower well depths and to the right to higher rod
concentration as length fractionation effects diminish. \ The critical value
of $U_{\min }^{\ast }$\ drops with decreasing $\sigma $.

Figure 5: \ Nematic cloud curve for $D_{1}/d=1$\ for various values of $%
\sigma $. \ The cloud curves increase to $\phi =1$\ as attractive
interactions increase, but this effect diminishes with decreasing $\sigma $%
.\ 

Figure 6: \ SWNT/superacid phase behavior as a function of SWNT volume
fraction and acid strength (measured by $dG$). \ Black symbols denote
experimental results \cite{davisscience}. Red symbols refer to theoretical
predictions, and red open circles ($\circ $) designate the theoretical
predictions for $\phi ^{(I)}$\ compared with experimental measurements of $%
\phi ^{(I)}$, denoted by black circles ($\bullet $). \ Black/red diamonds ($%
\diamond $) indicate the initial system concentration $\phi ^{(0)}$\ for
both experiments and simulations of isotropic\ (I)\ - liquid crystalline
(LC) phase separation. The open red triangles ($\bigtriangleup $) is the
Onsager predictions for $\phi ^{(I)}$\ for a system of monodisperse
hard-rods. \ \ A\ red line (--) represents the theoretical predictions for
the isotropic cloud curve. \ Simulations are for $\sigma =0.6$, $%
L_{mean}=348d$, $D_{1}/d=2.08$.

Figure 7: \ SWNT/120 H$_{2}$SO$_{4}$\ mixtures of varying $\phi ^{(0)}$\
were prepared and phase separated; the isotropic concentration $\phi ^{(0)}$%
\ was measured by UV-vis-nIR absorbance. \ These experimental measurements
showed a close match with theoretical predictions ($\sigma =0.6$, $%
L_{mean}=348d$) for phase separation as a function of $\phi ^{(0)}$.

Figure 8: \ (Left) Volumetric fraction of system that is nematic (denoted as 
$\gamma $) as a function of $\phi ^{(0)}$\ for for $U_{\min }=0$, $\sigma
=0.6$, $L_{mean}=348d$, $D_{1}/d=2.08$.\ The system reaches $\gamma =0.95$\
at $\phi ^{(0)}\approx 0.1$\ (noted as triangle), but the concentration must
increase to $\phi ^{(0)}=0.115$\ in order to reach $\gamma =1$\ (the nematic
cloud point, noted as diamond). \ Experimental data indicates a cloud point
of $\phi ^{(0)}\approx 0.1$\ but cannot detect the small isotropic regions
filled with short rods that persist up to $\phi ^{(0)}=0.115$. \ (Right)\
Average length of rods in the isotropic phase (denoted as $\langle
L_{I}\rangle $) as a function of $\phi ^{(0)}$. \ Near the isotropic cloud
point, $\langle L_{I}\rangle $\ approaches that of the parent population,
but $\langle L_{I}\rangle $\ decreases dramatically as $\phi ^{(0)}$\
approaches the nematic cloud point. \ Thus, near the nematic cloud point,
the rods in the isotropic phase tend to be the shortest rods in the parent
distribution, as confirmed experimentally elsewhere \cite{ZhangKinloch}.

\newif\ifabfull\abfulltrue

\end{document}